\documentclass[twocolumn]{aastex62}

\received{\ldots}
\revised{\ldots}
\accepted{\ldots}
\published{\ldots}
\reportnum{Accepted for publication in ApJL}

\shorttitle{A very young radio-loud magnetar}
\shortauthors{Esposito et al.}
\begin{document}

\title{A VERY YOUNG RADIO-LOUD MAGNETAR}
%% you desire. In v6.2 you can include a footnote in the title.

%% Use \email to set provide email addresses. Each \email will appear on its
%% own line so you can put multiple email address in one \email call. A new
%% \correspondingauthor command is available in V6.2 to identify the
%% corresponding author of the manuscript. It is the author's responsibility
%% to make sure this name is also in the author list.
%%

\correspondingauthor{Paolo Esposito}
\author[0000-0003-4849-5092]{P. Esposito}
\email{paolo.esposito@iusspavia.it}
\affil{Scuola Universitaria Superiore IUSS Pavia, Palazzo del Broletto, piazza della Vittoria 15, 27100 Pavia, Italy}
\affil{INAF--Istituto di Astrofisica Spaziale e Fisica Cosmica di Milano, via A.\,Corti 12, 20133 Milano, Italy}

\author[0000-0003-2177-6388]{N. Rea}
\affiliation{Institute of Space Sciences (ICE, CSIC), Campus UAB, Carrer de Can Magrans s/n, 08193, Barcelona, Spain}
\affiliation{Institut d'Estudis Espacials de Catalunya (IEEC), Carrer Gran Capit\`a 2--4, 08034 Barcelona, Spain}

\author[0000-0001-8785-5922]{A. Borghese}
\affiliation{Institute of Space Sciences (ICE, CSIC), Campus UAB, Carrer de Can Magrans s/n, 08193, Barcelona, Spain}
\affiliation{Institut d'Estudis Espacials de Catalunya (IEEC), Carrer Gran Capit\`a 2--4, 08034 Barcelona, Spain}

\author[0000-0001-7611-1581]{F. Coti Zelati}
\affiliation{Institute of Space Sciences (ICE, CSIC), Campus UAB, Carrer de Can Magrans s/n, 08193, Barcelona, Spain}
\affiliation{Institut d'Estudis Espacials de Catalunya (IEEC), Carrer Gran Capit\`a 2--4, 08034 Barcelona, Spain}

\author[0000-0001-7795-6850]{D. Vigan\`o}
\affiliation{Institute of Space Sciences (ICE, CSIC), Campus UAB, Carrer de Can Magrans s/n, 08193, Barcelona, Spain}
\affiliation{Institut d'Estudis Espacials de Catalunya (IEEC), Carrer Gran Capit\`a 2--4, 08034 Barcelona, Spain}
\author[0000-0001-5480-6438]{G. L. Israel}
\affil{INAF--Osservatorio Astronomico di Roma, via Frascati 33, 00078 Monteporzio Catone, Italy}
\author[0000-0002-6038-1090]{A. Tiengo}
\affil{Scuola Universitaria Superiore IUSS Pavia, Palazzo del Broletto, piazza della Vittoria 15, 27100 Pavia, Italy}
\affil{INAF--Istituto di Astrofisica Spaziale e Fisica Cosmica di Milano, via A.\,Corti 12, 20133 Milano, Italy}
\affil{Istituto Nazionale di Fisica Nucleare (INFN), Sezione di Pavia, via A.\,Bassi 6, 27100 Pavia, Italy}
\author[0000-0001-6762-2638]{A. Ridolfi}
\affil{INAF--Osservatorio Astronomico di Cagliari, Via della Scienza 5, 09047 Selargius, Italy}
\affil{Max-Planck-Institute f\"ur Radioastronomie, Auf dem Hugel 69, 53121 Bonn, Germany}
\author[0000-0001-5902-3731]{A. Possenti}
\affil{INAF--Osservatorio Astronomico di Cagliari, Via della Scienza 5, 09047 Selargius, Italy}
\affil{Department of Physics, Universit\`a di Cagliari, S.P. Monserrato-Sestu km 0,700, 09042 Monserrato, Italy}
\author[0000-0002-8265-4344]{M. Burgay}
\affil{INAF--Osservatorio Astronomico di Cagliari, Via della Scienza 5, 09047 Selargius, Italy}
\author{D. G\"otz}
\affil{AIM-CEA/DRF/Irfu/Service d’Astrophysique, Orme des Merisiers, F-91191 Gif-sur-Yvette, France}
\author[0000-0002-3869-2925]{F. Pintore}
\affil{INAF--Istituto di Astrofisica Spaziale e Fisica Cosmica di Milano, via A.\,Corti 12, 20133 Milano, Italy}
\author[0000-0002-0018-1687]{L. Stella}
\affil{INAF--Osservatorio Astronomico di Roma, via Frascati 33, 00078 Monteporzio Catone, Italy}
\author[0000-0003-0554-7286]{C. Dehman}
\affiliation{Institute of Space Sciences (ICE, CSIC), Campus UAB, Carrer de Can Magrans s/n, 08193, Barcelona, Spain}
\affiliation{Institut d'Estudis Espacials de Catalunya (IEEC), Carrer Gran Capit\`a 2--4, 08034 Barcelona, Spain}
\author{M. Ronchi}
\affiliation{Institute of Space Sciences (ICE, CSIC), Campus UAB, Carrer de Can Magrans s/n, 08193, Barcelona, Spain}
\affiliation{Institut d'Estudis Espacials de Catalunya (IEEC), Carrer Gran Capit\`a 2--4, 08034 Barcelona, Spain}
\author[0000-0001-6278-1576]{S. Campana}
\affil{INAF--Osservatorio Astronomico di Brera, via E.\,Bianchi 46, 23807 Merate (LC), Italy}
\author[0000-0002-9575-6403]{A. Garcia-Garcia}
\affiliation{Institute of Space Sciences (ICE, CSIC), Campus UAB, Carrer de Can Magrans s/n, 08193, Barcelona, Spain}
\affiliation{Institut d'Estudis Espacials de Catalunya (IEEC), Carrer Gran Capit\`a 2--4, 08034 Barcelona, Spain}
\author[0000-0002-6558-1681]{V. Graber}
\affiliation{Institute of Space Sciences (ICE, CSIC), Campus UAB, Carrer de Can Magrans s/n, 08193, Barcelona, Spain}
\affiliation{Institut d'Estudis Espacials de Catalunya (IEEC), Carrer Gran Capit\`a 2--4, 08034 Barcelona, Spain}
\author[0000-0003-3259-7801]{S. Mereghetti}
\affil{INAF--Istituto di Astrofisica Spaziale e Fisica Cosmica di Milano, via A.\,Corti 12, 20133 Milano, Italy}
\author[0000-0002-3635-5677]{R. Perna}
\affil{Department of Physics and Astronomy, Stony Brook University, Stony Brook, NY, 11794, USA}
\affil{Center for Computational Astrophysics, Flatiron Institute, 162 5th Avenue, New York, NY 10010, USA}
\author[0000-0003-3952-7291]{G.\,A. Rodr\'iguez Castillo}
\affil{INAF--Osservatorio Astronomico di Roma, via Frascati 33, 00078 Monteporzio Catone, Italy}
\author[0000-0002-5825-8149]{R. Turolla}
\affil{Dipartimento di Fisica e Astronomia `Galileo Galilei', Universit\`a di Padova, via F. Marzolo 8, 35131 Padova, Italy}
\affil{Mullard Space Science Laboratory, University College London, Holmbury St. Mary, Dorking, Surrey RH5 6NT, UK}
\author[0000-0001-5326-880X]{S. Zane}
\affil{Mullard Space Science Laboratory, University College London, Holmbury St. Mary, Dorking, Surrey RH5 6NT, UK}

\def\xmm {\emph{XMM--Newton}}
\def\cxo {\emph{Chandra}}
\def\nustar {\emph{NuSTAR}}
\def\rst {\emph{ROSAT}}
\def\swift {\emph{Swift}}
\def\nicer {\emph{NICER}}
\def\src {\mbox{Swift\,J1818}}

\def\flux {\mbox{erg cm$^{-2}$ s$^{-1}$}}
\def\lum {\mbox{erg s$^{-1}$}}
\def\nh {$N_{\rm H}$}
\def\kms  {\rm \ km \, s^{-1}}
\def\cms  {\rm \ cm \, s^{-1}}
\def\gs   {\rm \ g  \, s^{-1}}
\def\cmtre {\rm \ cm^{-3}}
\def\cmdue {\rm \ cm^{-2}}
\def\ss {\mbox{s\,s$^{-1}$}}
\def\chisq {$\chi ^{2}$}
\def\rchisq {$\chi_{\nu} ^{2}$}

\def\arc{\mbox{$^{\prime\prime}$}}
\def\arcmin{\mbox{$^{\prime}$}}
\def\deg{\mbox{$^{\circ}$}}

\def\rsun {~R_{\odot}}
\def\msun {~M_{\odot}}
\def\mdotav {\langle \dot {M}\rangle }

\def\uu {4U\,0142$+$614}
\def\ee {1E\,1048.1$-$5937}
\def\kes {1E\,1841$-$045}
\def\aa {1E\,1547$-$5408}
\def\axj {AX\,J1844$-$0258}
\def\rxs {1RXS\,J1708$-$4009}
\def\xte{XTE\,J1810$-$197}
\def\smc{CXOU\,J0100$-$7211\,}
\def\wes{CXOU\,J1647$-$4552}
\def\ea {1E\,2259$+$586}
\def\sgra{SGR\,1806$-$20}
\def\sgrb{SGR\,1900$+$14}
\def\sgrd{SGR\,1627$-$41}
\def\sgre{SGR\,0501$+$4516}
\def\sgrf{SGR\,1935+2154}
\def\lowba{SGR\,0418$+$5729}
\def\sgrg{SGR\,1833$-$0832}
\def\lowbb{Swift\,J1822.3$-$1606}
\def\galmag{PSR\,J1745$-$2900}
\def\sgras{Sgr\,A$^{\star}$}
\def\sgrh{SGR\,1801$-$21}
\def\sgri{SGR\,2013$+$34}
\def\psr{PSR\,1622$-$4950}
\def\hbpsr{PSR\,J1846$-$0258}
\def\radiohb{PSR\,J1119$-$6127}

\begin{abstract}
The magnetar Swift\,J1818.0--1607 was discovered in March 2020  when  \emph{Swift} detected a 9\,ms hard X-ray burst and a long-lived outburst. Prompt X-ray observations revealed a spin period of $1.36$\,s, soon confirmed by the discovery of radio pulsations. We report here on the analysis of the \emph{Swift} burst and  follow-up X-ray and radio observations. The burst average luminosity was  $L_{\rm burst} \sim2\times 10^{39}$\,\lum\ (at 4.8\,kpc). Simultaneous observations with \xmm\ and \nustar\ three days after the burst provided a source spectrum well fit by an absorbed blackbody (\nh =  $(1.13\pm0.03) \times 10^{23}$\,cm$^{-2}$ and $kT = 1.16\pm0.03$\,keV) plus a power-law ($\Gamma=0.0\pm1.3$) in the 1--20\,keV band, with a luminosity of  $\sim$$8\times10^{34}$ \lum, dominated by the blackbody emission. From our timing analysis, we derive a dipolar magnetic field $B \sim 7\times10^{14}$\,G,  spin-down luminosity $\dot{E}_{\rm rot} \sim 1.4\times10^{36}$\,\lum\  and characteristic age of 240~yr, the shortest currently known. Archival observations led to an upper limit on the quiescent luminosity $<$$5.5\times10^{33}$\,\lum, lower than the value expected from magnetar cooling models at the source characteristic age.  A 1\,hr radio observation with the Sardinia Radio Telescope taken about 1 week after the X-ray burst detected a number of strong and short radio pulses at 1.5\,GHz, in addition to regular pulsed emission; they were emitted at an average rate 0.9\,min$^{-1}$ and accounted for $\sim$50\% of the total pulsed radio fluence. We conclude that Swift\,J1818.0--1607 is a peculiar magnetar belonging to the small, diverse group of young neutron stars with properties straddling those of rotationally and magnetically powered pulsars. Future observations will make a better estimation of the age possible by measuring the spin-down rate in quiescence.
%A patch of diffuse emission surrounding the source, likely a scattering halo, was found.
%A \emph{Swift} monitoring campaign  caught Swift\,J1818.0--1607 at steady X-ray flux over the subsequent two weeks. 
%This hints to an emission mechanism resembling the case of the ``Giant Pulses" from radio pulsars. 
%Simulations of the source magneto-thermal evolution predict a larger magnetar age and quiescent luminosity than inferred from our observations.
\end{abstract}
%\keywords{Magnetars(992) --- Neutron stars(1108) --- Radio pulsars(1353) --- Transient sources(1851) --- X-ray bursts(1814)}

\section{Introduction} \label{sec:intro}

 The emission of magnetars is believed to be powered by the dissipation of their unstable strong magnetic fields ($B\sim 10^{14}$--$10^{15}$\,G; see \citealt{kaspi17,esposito18} for recent reviews). 
 At variance, in the vast majority of radio pulsars, rotational energy provides the energy budget for particle acceleration, ultimately leading to their radio to gamma-ray emission. 
However, a well-defined dichotomy between the two classes was shown to be inadequate.
In particular, magnetar-like X-ray activity was found from pulsars with powerful rotational energy loss rate, such as \hbpsr\ \citep{gavriil08} and \radiohb\ \citep{archibald16}, whereas pulsed radio emission was detected from several magnetars in outburst. Moreover, enigmatic magnetars having dipolar magnetic fields as low as a few $10^{12}$\,G \citep{rea10,rea13} or spin periods of the order of a few hours \citep{deluca06,rea16} were discovered.
These findings hint at a complex, more
compounded picture.

On 2020 March 12, the Burst Alert Telescope (BAT) on board the \emph{Neil Gehrels Swift Observatory} \citep{gehrels04} triggered on a burst, which was soon recognized to have characteristics typical of those of short bursts from magnetars \citep{evans20}. The \swift\ X-ray Telescope (XRT) started to observe the field about 64\,s afterwards, and detected a new uncatalogued X-ray source, Swift\,J1818.0--1607 (henceforth dubbed \src). Follow-up observations with \nicer\ detected a coherent periodic X-ray signal at 1.36\,s \citep{enoto20}. Furthermore, radio observations from several antennas identifed \src\ as the fifth radio-loud magnetar \citep{karuppusamy20} and provided a first measurement of the spin period derivative of $8.2\times10^{-11}$\,s\,s$^{-1}$ \citep{champion20}, converting to a dipolar magnetic field of $B\sim 6.8\times 10^{14}$\,G and a very small characteristic age $<$300\,yr. 

This Letter reports on: i) the burst detected by \swift/BAT that led to the discovery of \src, ii) prompt simultaneous X-ray observations using \xmm\ and \nustar, iii) the \swift/XRT monitoring campaign over the first three weeks of the outburst, iv) radio observations with the Sardinia Radio Telescope (SRT) in the P (0.34\,GHz) and L (1.5\,GHz) bands, performed one week after the burst detection (\S\ref{sec:X} and \S\ref{section:srt}). Summary of the results and discussion follow (\S\ref{discussion}).

\section{X-ray emission}
\label{sec:X}
\subsection{Observations and data analysis} 

%%%%%%%%%%%%%%%%%%%%%%%%%%%%%%%%%%%%%%%%%%%%%%%%%%%%%%%%%%%%%%%%%%%%%%%%
% START TABLE
\begin{deluxetable*}{cccccc}[htb]
\tablecaption{Observation log.
\label{tab:observations}}
\tablecolumns{5}
\tablenum{1}
\tablewidth{0pt}
\tablehead{
\colhead{Instrument\tablenotemark{a}} &
\colhead{Obs.ID} &
\colhead{Start} &
\colhead{Stop} & 
\colhead{Exposure} & \colhead{Count rate\tablenotemark{b}}\\
\colhead{} &
\colhead{} & 
\multicolumn2c{YYYY-MM-DD hh:mm:ss (TT)} &
\colhead{(ks)} & \colhead{(counts s$^{-1}$)}
}
\startdata
\rst/PSPC  & 50311 & 1993-09-12 20:55:04 & 1993-09-13 18:09:28 & 6.7 & $<$$10^{-3}$\\
\cxo/ACIS-I (TE) & 8160 & 2008-02-16 06:46:59 & 2008-02-16 07:51:04 & 2.7 & $<$$2.5\times10^{-3}$\\
\emph{XMM}/EPIC-pn (FF) & 0152834501 & 2003-03-28 04:35:03 & 2003-03-28 07:26:37 & 8.4 & $<$0.062\\
\swift/XRT (PC) & 00032293013 & 2012-03-17 15:03:55 & 2012-03-17 15:25:56 & 1.3 & $<$0.026 \\
\swift/XRT (PC) & 00044110002 & 2012-10-12 17:14:05 & 2012-10-12 17:16:55 & 0.2 & $<$0.089 \\
%\swift/XRT (PC) & 00044112002 & 2012-10-17 00:09:18 & 2012-10-17 00:15:54 & 0.4 & $<$0.029(?) \\
\swift/XRT (PC) & 00044110003 & 2012-10-17 13:08:36 & 2012-10-17 13:11:55 & 0.2 & $<$0.113 \\
\swift/XRT (PC) & 00044111003 & 2012-10-23 21:07:27 & 2012-10-23 21:12:56 & 0.3 & $<$0.043 \\
\swift/XRT (PC) & 00087426001 & 2017-07-24 19:11:45 & 2017-07-24 21:00:52 & 2.2 & $<$$8.1\times10^{-3}$ \\
\swift/XRT (PC) & 00087426002 & 2017-07-27 21:45:57 & 2017-07-27 23:37:52 & 2.7 & $<$$4.9\times10^{-3}$ \\
\emph{XMM}/EPIC-pn (FF) & 0800910101 & 2018-04-08 21:27:40 & 2018-04-09 14:24:19 & 60.4 & $<$$8.3\times10^{-3}$\\
\hline
\swift/XRT (PC) & 00960986000 & 2020-03-12 21:18:22 & 2020-03-12 21:36:48 & 1.1 & 0.15 $\pm$ 0.01 \\
\swift/XRT (PC) & 00960986001 & 2020-03-12 22:57:45 & 2020-03-13 05:13:02 & 4.9 &  0.14 $\pm$ 0.01 \\
\swift/XRT (WT) & 00960986002 & 2020-03-13 20:47:55 & 2020-03-13 21:21:15 & 2.0 &  0.16 $\pm$ 0.01 \\
\swift/XRT (PC) & 00960986003 & 2020-03-15 00:10:37 & 2020-03-15 03:36:52 & 1.5 &  0.14 $\pm$ 0.01 \\
\nustar/FPMA & 80402308002 & 2020-03-15 03:58:21 & 2020-03-15 15:58:03 & 22.2 & $0.443\pm0.005$  \\
%\emph{XMM}/MOS1 (SW) & 0823591801 & 2020-03-15 07:11:08  & 2020-03-15 14:30:33  & 25.6 \\
%\emph{XMM}/MOS2 (SW) & 0823591801 & 2020-03-15 07:11:28 & 2020-03-15 14:33:48  & 25.7 \\
\emph{XMM}/EPIC-pn (LW)  & 0823591801 & 2020-03-15 07:57:47 & 2020-03-15 14:41:12  & 22.1 & $1.45\pm0.01$ \\
\swift/XRT (WT) & 00960986004 & 2020-03-19 09:33:11 & 2020-03-19 11:16:56 & 1.7 & 0.19 $\pm$ 0.02 \\
\swift/XRT (WT) & 00960986005 & 2020-03-20 04:34:19 & 2020-03-20 04:49:56 & 1.8 & 0.20 $\pm$ 0.01 \\
\swift/XRT (WT) & 00960986006 & 2020-03-22 02:35:21 & 2020-03-22 03:01:56 & 1.6 & 0.16 $\pm$ 0.01 \\
\swift/XRT (WT) & 00960986007 & 2020-03-24 05:51:38 & 2020-03-24 09:02:56 & 1.2 & 0.13 $\pm$ 0.01 \\
\swift/XRT (WT) & 00960986008 & 2020-03-26 05:40:29 & 2020-03-26 23:20:56 & 1.1 & 0.19 $\pm$ 0.01 \\
\swift/XRT (WT) & 00960986009 & 2020-03-28 03:40:53 & 2020-03-28 18:07:56 & 1.2 & 0.18 $\pm$ 0.02 \\
\swift/XRT (WT) & 00960986010 & 2020-03-29 16:25:13 & 2020-03-30 21:03:56 & 1.3 & 0.16 $\pm$ 0.01 \\
\swift/XRT (WT) & 00960986011 & 2020-04-01 19:17:34 & 2020-04-01 19:25:56 & 0.5 & 0.17 $\pm$ 0.02 \\
\enddata
\tablenotetext{a}{The instrumental setup is indicated in brackets: TE = timed exposure}, FF = full frame, PC = photon counting, WT = windowed timing, and LW = large window.
\tablenotetext{b}{The count rate is in the 0.3--10\,keV energy range, except for \rst\ (0.1--2.4\,keV) and \nustar\ (3--20\,keV); if the source was not detected, we give a 3$\sigma$ upper limit.}
%\tablecomments{} SW = small window,
\end{deluxetable*}
%%%%%%%%%%%%%%%%%%%%%%%%%%%%%%%%%%%%%%%%%%%%%%%%%%%%%%%%%%%%%%%%%%%%%%%%

\subsubsection{Swift}\label{sec:swift}

After the \swift/BAT detection of the burst and the prompt slew of the spacecraft (obs.ID 00960986000), several \swift/XRT observations of \src\ were carried out, in both photon counting (PC, CCD readout time of 2.5\,s) and windowed timing (WT, readout time of 1.8\,ms) modes (see Table\,\ref{tab:observations}).
The data were processed and analyzed using standard procedures and software packages (\textsc{heasoft} v.\,6.25, \textsc{caldb} 2020-01-09). The source photons were selected within a 20-pixel radius (1 pixel $ = 2\farcs36$). \swift/BAT mask-tagged light curves, images and spectra were created only for the burst event.

\subsubsection{XMM--Newton}\label{sec:xmm}

\src\ was observed with the European Photon Imaging Camera (EPIC) on board the \xmm\ satellite on 2020 March 15 for an on-source exposure time of 22.1\,ks (Table\,\ref{tab:observations}). The EPIC-pn \citep{struder01} was set in large window mode (LW; timing resolution of 47.7\,ms), while both MOS detectors \citep{turner01} were operating in small window (SW; timing resolution of 0.3\,s) mode.  In this Letter, we use only the data acquired with the EPIC-pn camera, owing to its higher time resolution and better capability to model diffuse emission around the source (\S\ref{xrayanalysis}) compared to the central CCD of the MOS operated in SW. %, which provides higher counting statistics than the MOS detectors.}
%All cameras mounted the thin optical-blocking filter. 
 The raw data were analyzed with the {\sc SAS} v.\,18.0.0 software package. We cleaned the observations from periods of high background activity; in the EPIC-pn, this resulted in a net exposure of 14\,ks. 
 %We cleaned the observations from particle flares using good-time intervals created from light curves of the entire field of view above 10\,keV and employing an intensity threshold. 

\begin{figure}
\centering
\resizebox{\hsize}{!}{\includegraphics[angle=0]{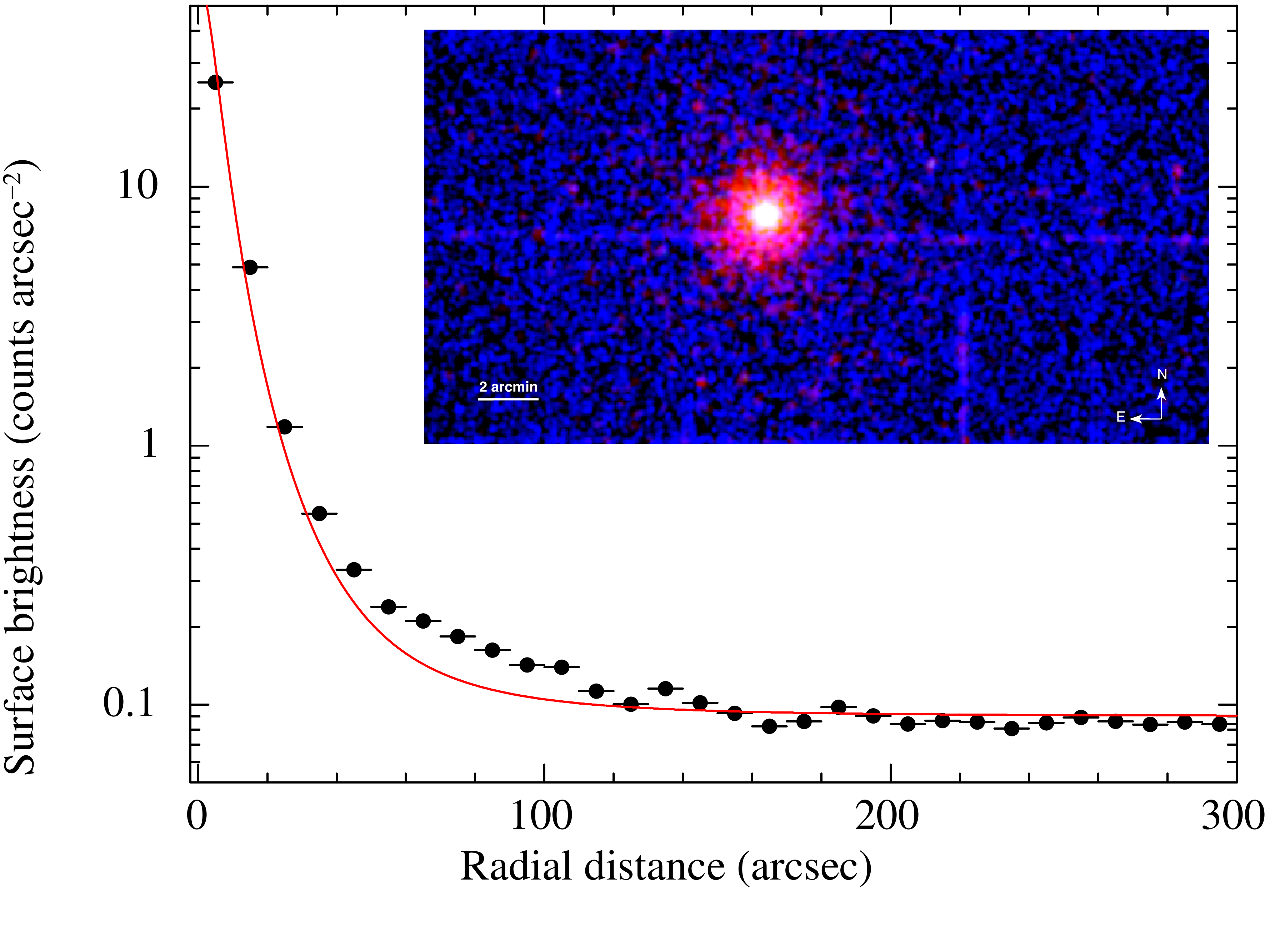}}
\vspace{-0.5cm}
\caption{\label{diffuse} Observed X-ray surface brightness up to a radial distance of 300\,arcsec from the source in the 0.3--10\,keV energy range (note that the error bars are smaller than the symbols). The red curve represents the PSF model. The inset shows a false-color X-ray image from the EPIC-pn observation. Red, green and blue colors correspond to the 2--4, 4--7.5, and 8.5--12\,keV energy bands, respectively.}
\end{figure}

We detected diffuse emission around the source (Fig.\,\ref{diffuse}). To quantify its spatial extension, we extracted the radial profile of the observed surface brightness up to a radial distance of 300\,arcsec from the source. We then modelled it using a King function reproducing the  EPIC-pn point-spread function \citep{ghizzardi02} plus a constant term accounting for the background level. A photon excess associated with the diffuse emission is present at radial distances within the  $\approx$50--110\,arcsec range (Fig.\,\ref{diffuse}). We selected the source photon counts from a circle of 40\,arcsec radius, and those of the diffuse emission from an annulus with radii of 50 and 110\,arcsec. The background level was estimated using a 100-arcsec circle far from the source, on the same CCD.
The average background-subtracted surface brightness of the diffuse emission was $(0.086\pm0.002)$\,counts arcsec$^{-2}$ (0.3--10\,keV).
%$(8.3\pm0.2)\times10^{-6}$\,counts s$^{-1}$ arcsec$^{-2}$ (0.3--10\,keV)
% The response matrix and ancillary files for the spectrum of the source and the diffuse emission were generated through the {\sc rmfgen} and {\sc arfgen} tools, respectively.

%%%%%%%%%%%%%%%%%%%%%%%%%%%%%%%%%%%%%%%%%%%%%%%%%%%%%%%%%%%%%%%
%%%%%%%%%%%%%%%%%%%%%%%%%%%%%%%%%%%%%%%%%%%%%%%%%%%%%%%%%%%%%%%

\subsubsection{NuSTAR}\label{sec:nustar}

\nustar\ \citep{harrison13} observed \src\ on 2020 March 15 for an on-source exposure time of 22.2\,ks (Table\,\ref{tab:observations}). 
We reprocessed the event lists and filtered out passages of the satellite through the South Atlantic Anomaly using the {\sc nupipeline} script in the {\sc nustardas} 1.9.3 package with the latest calibration files (v.\,20200429). Stray-light contamination from a source outside the field of view is evident for both modules, but particularly strong in the FPMB data.  \src\ was detected up to $\sim$20\,keV and $\sim$15\,keV in the FPMA and FPMB, respectively. A circle with a radius of 100 arcsec was used to collect source photons ($\sim$90\% enclosed energy fraction; \citealt{madsen15}), while background counts were extracted from a 100-arcsec circle located on the same chip as the target.
%(the signal-to-noise ratio, S/N, decreases when considering broader energy bands). 
To study the source emission up to the highest energies we used only the data from the FPMA. We extracted light curves and spectra and generated instrumental response files using {\sc nuproducts}.  
% To study the source emission over the broadest energy range available from our observation, we used only the FPMA data in the spectral analysis.

\subsection{Results of the X-ray analysis}
\label{results}

\subsubsection{Burst Properties}
\begin{figure*}
\centering
%\resizebox{\hsize}{!}{\includegraphics[angle=0, height=10cm]{lcurves_profiles_spectra.pdf}}
\resizebox{\hsize}{!}{\includegraphics[angle=0,height=10cm]{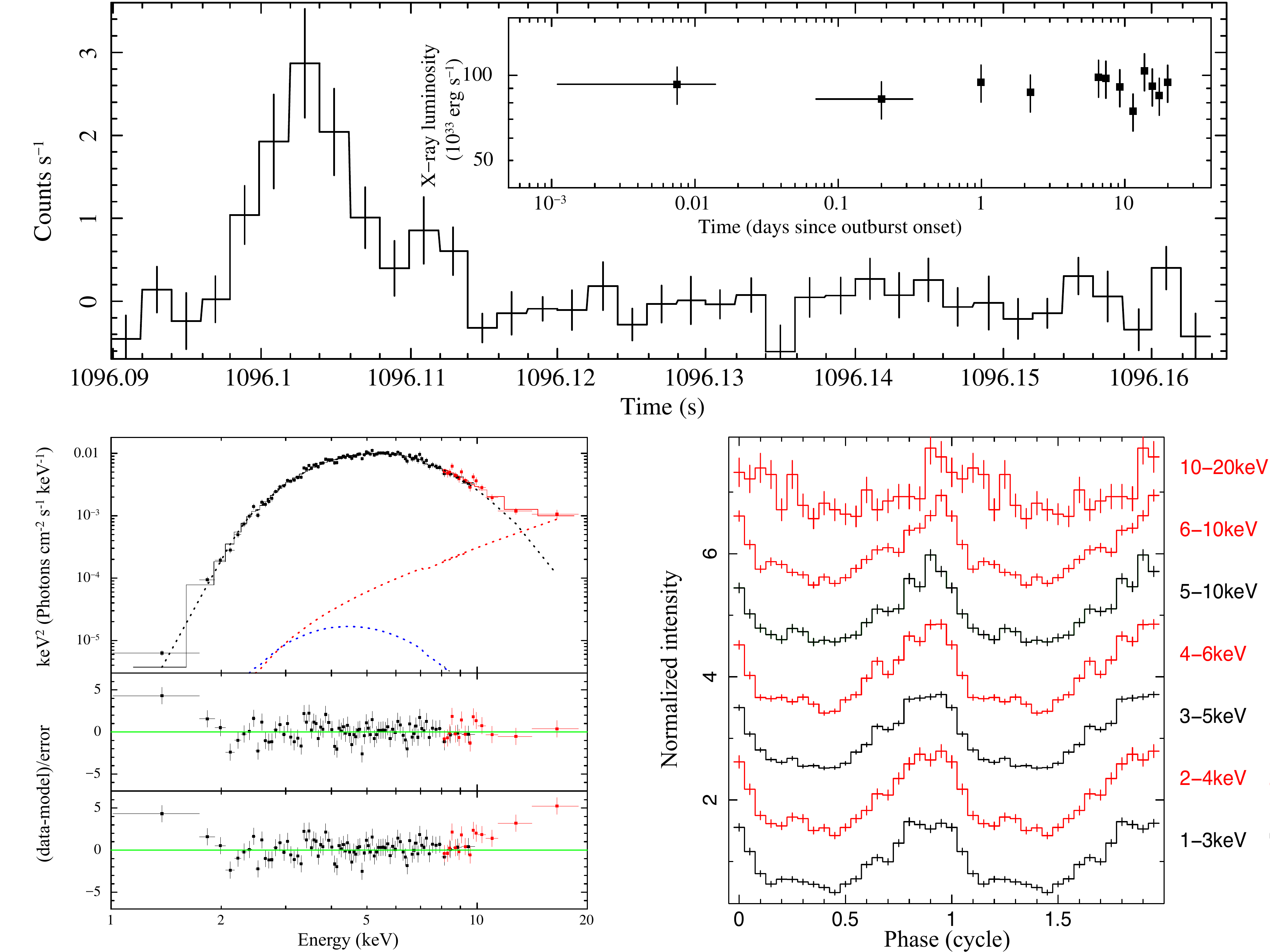}}
\caption{\label{spectrum} \emph{Top:} BAT light curve (15--150\,keV, bin time: 2\,ms; the start time is arbitrary). The inset shows the  evolution of the 0.3--10\,keV luminosity of \src\ measured with \swift/XRT since the burst trigger (MJD 58920.8867). \emph{Bottom-left panel, top}: Broad-band unfolded spectrum extracted from the simultaneous \xmm/EPIC-pn (in black) and \nustar/FPMA (red) data. The solid line shows the best-fitting model. The blue, black and red dotted lines indicate the diffuse emission component, the source blackbody and power law components, respectively. \emph{Bottom-left panel, middle}: Post-fit residuals. \emph{Bottom-left panel, bottom}: Residuals after removing the power law component from the model. \emph{Bottom-right panel}: Energy-resolved pulse profiles of \src\ extracted from the EPIC-pn (in black) and \nustar/FPMA (red) data.}
\end{figure*}

The burst had a T90 duration (the time interval containing 90\% of the counts) of $8\pm2$\,ms and a total duration of $\sim$9\,ms. These values were computed by the Bayesian blocks algorithm \textsc{battblocks} on mask-weighted light curves binned at 1\,ms in the 15--150\,keV range (the light curve of the event is shown in Fig.\,\ref{spectrum}), where essentially all the emission is contained.
%%%%%%%%%%%%%%%%%%%%%%%%%%%%%%%%%%%%%%%%%%%%%%%%%%%%%%%%%%%%%%%%%%%%%%%%
%%%%%%%%%%%%%%%%%%%%%%%%%%%%%%%%%%%%%%%%%%%%%%%%%%%%%%%%%%%%%%%%%%%%%%%%

We tested a blackbody, a power law, and an optically-thin thermal bremsstrahlung to the time-averaged spectrum.
%extracted from the total duration of 9\,ms with. 
All models provided good fits, with reduced $\chi^2$, $\chi^2_\nu = 0.62$ for 56 degrees of freedom (dof) in the case of a blackbody
 (temperature of $kT=6.4\pm0.7$\,keV), $\chi^2_\nu = 0.70$/56\,dof for the power law (photon index $\Gamma=3.1^{+0.3}_{-0.2}$), and $\chi^2_\nu = 0.65$/56\,dof for the bremsstrahlung ($kT=21^{+5}_{-4}$\,keV). From the blackbody fit we found an average flux $(6.2\pm0.9)\times10^{-7}$~\flux\ in the  15--150\,keV range, corresponding to an isotropic luminosity of $(1.7\pm0.3)\times10^{39}d^2_{4.8}$\,\lum, where $d_{4.8}$ is the source distance in units of 4.8\,kpc \citep{karuppusamy20}.

\subsubsection{Persistent emission}\label{xrayanalysis}

The \xmm/EPIC-pn and \nustar/FPMA background-subtracted spectra were grouped so as to have at least 100 and 50 counts per bin, respectively.
The spectral analysis was performed with {\sc xspec}. Absorption by the interstellar medium was modelled using the {\sc TBabs} model with the abundances from \citet{wilms00}.
% We extracted the spectra between 0.3 and 10\,keV but, after inspecting the data, we limited the analysis to the 1--10\,keV energy range, because of the extremely low signal-to-noise ratio (S/N) of \src\ below 1\,keV.
We extracted the spectra in the 0.3--10\,keV range but, after inspecting the data, we limited the analysis to the 1--10 keV energy range, because of the  very low signal-to-noise ratio (S/N) of \src\ below 1\,keV.

Firstly, we fit the EPIC-pn spectrum of the diffuse emission with an absorbed blackbody model, deriving a temperature $kT_{\rm diff} = (0.88\pm0.02)$\,keV.
%Firstly, we fit an absorbed blackbody model to the EPIC-pn spectrum of the diffuse emission over the \mbox{2--10\,keV} range (a detailed study of this component will be presented in a forthcoming paper), yielding a temperature $kT_{\rm diff} = (0.88\pm0.02)$\,keV. 
We then modelled the \src\  EPIC-pn spectrum with two absorbed blackbody components, by fixing the first temperature to $kT_{\rm diff}$ and leaving all normalizations free to vary. We derived a column density of \nh\ = $(1.12 \pm 0.03) \times 10^{23}$\,cm$^{-2}$, a source temperature of $kT_{\rm BB} = (1.17\pm0.03)$\,keV and an emitting radius of $R_{\rm BB} = (0.57\pm0.02)$\,km (at 4.8\,kpc). The observed fluxes of the source and diffuse blackbody components in the 1--10\,keV range were $F_{\rm BB}\sim$1.4$\times 10^{-11}$\,\flux, and $F_{\rm diff}\sim$4$\times 10^{-13}$\,\flux, respectively. At a distance of 4.8\,kpc this corresponds to a source luminosity of $\sim$7$\times 10^{34}d^2_{4.8}$\,\lum.  

We then performed a joint fit of the EPIC-pn and FPMA spectra using the above-mentioned model plus a power-law component to model the source high-energy emission. We removed the FPMA data below 8\,keV to minimize contamination from diffuse emission at lower energies. All parameters were tied between the two spectra. The best-fitting values resulted: \nh\ = $(1.13 \pm 0.03) \times 10^{23}$\,cm$^{-2}$, $kT = (1.16\pm0.03)$\,keV, $R_{\rm BB} = (0.58\pm0.03)$\,km (at 4.8\,kpc) and photon index $\Gamma=0.0\pm1.3$ ($\chi^2_{\nu}=1.1$ for 109 dof; see Fig.\,\ref{spectrum}).
%The best-fitting values for the source blackbody component were consistent within the uncertainties with those reported above from the analysis of the EPIC-pn data alone. The photon index was $\Gamma=-0.02_{-1.38}^{+1.26}$ ($\chi^2_{\nu}=1.1$ for 109 dof). 
The total observed flux after subtracting the contribution from the diffuse emission was $\sim$$1.5\times 10^{-11}$\,\flux\ (1--20\,keV), giving a luminosity of $\sim8\times10^{34}d^2_{4.8}$\,\lum\ (1--20\,keV). 
The observed flux of the blackbody component was $\sim$$1.3\times 10^{-11}$\,\flux\ (1--20\,keV), accounting for most of the observed X-ray emission, and corresponding to a luminosity of $\sim$$6.8 \times 10^{34}d^2_{4.8}$\,\lum\ (1--20\,keV).
%For the power law, we found a 1-$\sigma$ upper limit on the observed flux of 1.4 $\times$ 10$^{-12}$\,\flux (0.3--20)\,keV, corresponding to a luminosity of $<5.3 \times 10^{33}d^2_{4.8}$\,\lum. 

Given the short exposure, poor statistics and S/N of the \swift/XRT observations, their analysis was mainly aimed at sampling the long-term flux evolution of \src, and supplementing the \xmm\ and \nustar\ timing analysis. For this reason, we fit simultaneously all the \swift\ spectra with an absorbed blackbody model (\nh\ was kept frozen at the above-mentioned value). 
%derived from the joint fit of the pn and FPMA spectra. 
%Once the best fit was found, the observed and unabsorbed fluxes were estimated with the convolution model {\sc cflux}. 
Fig.\,\ref{spectrum} shows the long-term light curve of \src.
From the \xmm\ spectral analysis, we estimate that the systematic uncertainty of fluxes and luminosities resulting from contamination by the diffuse emission is $\lesssim$15\,\% (if steady in time).
%where luminosities were calculated assuming a distance of 4.8\,kpc.
%together with the light curves of the most powerful outbursts detected so far that were intensively monitored during the first month since the onset.

%Due to the proximity of the magnetar candidate AX\,J1818.8--1559 \citep{mereghetti12} 
The field of \src\ was observed several times with sensitive imaging instruments before March 2020 (Table\,\ref{tab:observations}; \citealt{mereghetti12}). The source was not detected in any observation (see Table\,\ref{tab:observations} for the 3\,$\sigma$ upper limits on the count rate). The \cxo\ and the 2018 \xmm\ observations provided the deepest limits. Using the {\sc webpimms} tool\footnote{See \url{https://heasarc.gsfc.nasa.gov/cgi-bin/Tools/w3pimms/w3pimms.pl}.} and assuming an absorbed blackbody with $kT=0.3$\,keV and $N_{\rm H}=1. 2\times10^{23}$\,cm$^{-2}$, both their limits translate into a 0.3--10\,keV flux of $< 3.4\times10^{-14}$\,\flux, corresponding to a luminosity of $< 5.5\times10^{33}d_{4.8}^2$\,\lum. We also note that in the 2018 \xmm\ observation, the diffuse emission was not detectable, with an upper limit implying an emission at least $\sim$10 times fainter (Tiengo et al., in preparation).

\subsubsection{Timing Analysis}\label{section:timing}

For the timing analysis, we referred the photon arrival times to the Solar System barycenter using our best \swift\ position ($\rm RA=18^h18^m00\fs16$, $\rm Dec= -16^\circ07'53\farcs2$, J2000.0; uncertainty of 2\,arcsec at 90\% c.l.). %and the DE-200 Solar System ephemeris. 
By a phase-fitting analysis of the X-ray data, we measured a period $P = 1.363489(3)$\,s and a period derivative $\dot{P}=9(1)\times10^{-11}$\,s\,s$^{-1}$ (with epoch MJD\,58922.31 and valid over the MJD range \mbox{58923.5--58928.5}), compatible with previous radio timing measurements reported by \citet{champion20}. 
The energy-resolved pulse profiles extracted from EPIC-pn and \nustar\ data are shown in Fig.\,\ref{spectrum}. The background-subtracted peak-to-peak semi-amplitude increases with energy from ($52\pm2$)\,\% to ($66\pm2$)\,\% over the 1--10\,keV band (as measured with the EPIC-pn), and is equal to ($58\pm13$)\,\% in the 10--20\,keV range. The latter values are not corrected for the underlying diffuse emission component, which should affect the pulsed fraction values by a few percents. 

\section{Radio emission}\label{section:srt}

We observed \src\ with the SRT on 2020 March 19 at 05:05 UTC for 1\,h, using the coaxial L/P band receiver to observe simultaneously in two frequency bands, centered at 1548\,MHz and 336\,MHz, respectively.   
%The source was observed for 1 hour, starting on 2020 March 19 at 05:05 UTC.
In the L band, we recorded the total intensity signal in incoherent search mode over a usable bandwidth of $\sim$390\,MHz with frequency resolution of 1\,MHz and  time resolution of 100\,$\mu$s.  We de-dispersed and folded the data using our position of the source and the spin parameters and dispersion measure (DM) by \citet{champion20}. We extracted topocentric times-of-arrival and used them to determine the ${\rm DM}\,=700.8(6)$\,pc\,cm$^{-3}$, the spin frequency $\nu=0.7333920(2)$\,Hz ($P=1.3635273(4)$\,s, compatible within 1-$\sigma$ uncertainty with the measurement of \citealt{champion20}) and the pulse profile width (at 50\% of the peak) $W50\sim 40$\,ms at epoch MJD\,58927.23. 
The optimized $\rm S/N\sim22$ of the SRT observation implies an average flux density $S_{\rm ave}\sim 0.2$\,mJy, assuming an antenna gain $0.55$\,K/Jy and system temperature $\sim$30\,K during the observation ($S_{\rm ave}$ and all the energetic/fluence calculation below must be considered as a lower limit since the residual in-band RFI can have a significant impact on the value of the rms noise).

\begin{figure}
\centering
\resizebox{\hsize}{!}{\includegraphics[angle=0]{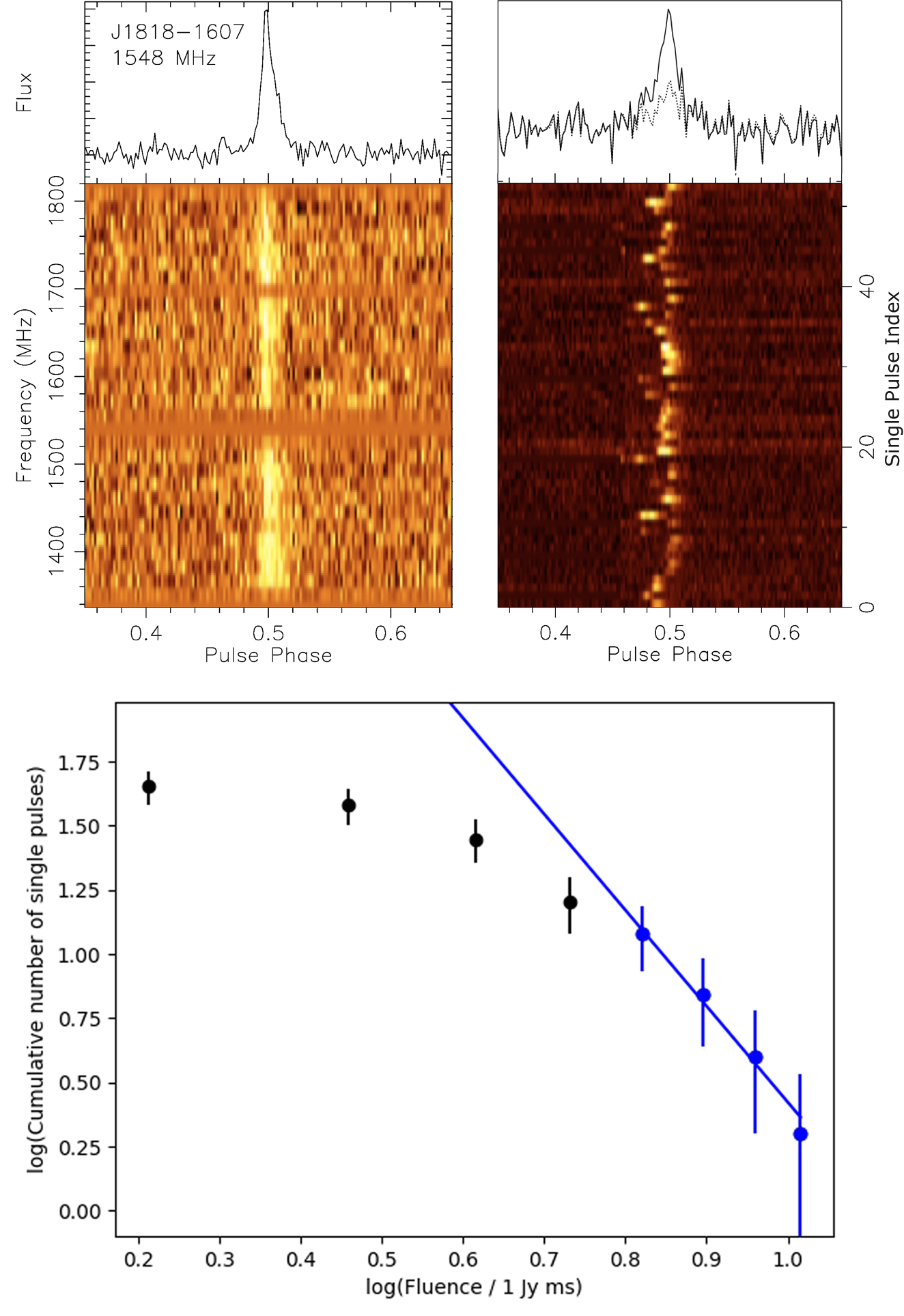}}
\vspace{-0.5cm}
\caption{\emph{Upper-left}:  Profile of the strongest single pulse detected with the SRT  (the flux is in arbitrary units) de-dispersed at a DM = 700.8\,pc\,cm$^{-3}$ (top) and waterfall plot (below) displaying the observing frequency (in $30 \times 16$-MHz-wide sub-bands) versus the pulse phase (which is arbitrary with respect to the X-ray pulse phase of Fig.\,\ref{spectrum}) for the same single pulse as the panel above. \emph{Upper-right}: The top panel shows the integrated pulse profile over the entire observation, lasting about 1\,hr (solid line), as well as the integrated profile after removing from the profile above the 53 strong single pulses mentioned in the text (dotted line). The two profiles of the upper panel are reported with the same phase reference (also identical to that of the left panels) of the detected series of the 53 single pulses, which are plotted below, on a colour scale, on top of each other. Although not arriving at a constant rotational phase, the latter are basically confined within the phase range of the total integrated pulse profile.
%The total duration of the 53 non consecutive reported rotations is about 72\,s. 
\emph{Lower panel}: Cumulative number of single pulses exceeding a given fluence. The uncertainties are considered to be those of a Poisson process. The line shows the power law fit to the blue points.
}
\label{fig:radio}
\end{figure}

A search for single pulses was carried out with {\sc PRESTO}\footnote{\url{http://www.cv.nrao.edu/~sransom/presto/}} \citep{ransom2001}. The data were down-sampled by a factor of four and de-dispersed at the above DM. The script {\sc single\_pulse\_search.py} was run with S/N threshold of 8 and maximum width of 0.1\,s, unveiling 53 pulses (Fig.\,\ref{fig:radio}). Their widths $(W50_{\rm SP})$ range from $\sim$7 to $\sim$22\,ms (significantly smaller than the width of the integrated profile) and their {\sc pdmp}\footnote{\url{http://psrchive.sourceforge.net}} S/N's range from 7 to 37. 
% No pulse sub-structure is recognizable down to the time resolution of our 1024-bin profiles (i.e. $\sim$1.3\,ms per bin), at odds with the multiple peaks seen at higher frequencies by \citet{gajjar20}. 
The energetic per rotation $E_{\rm SP}$ (the pulse-integrated flux density using a pulse width at 50\% of the peak and averaged over a spin period) of each single pulse was determined from the values above and compared with the average energetic per rotation of the radio emission $E_{\rm no-SP}$, after removing the rotations containing the aforementioned single pulses. We found a ratio $R_{\rm E}=E_{\rm SP}/E_{\rm no-SP}$ ranging from 16 to 126.

%%%%%%%%%%%%%%%%%%%%%%%%%%%%%%%%%%%%%%%%%%%%%%%%%%%%%%%%%%%%%%%%%%%%%%%%
%%%%%%%%%%%%%%%%%%%%%%%%%%%%%%%%%%%%%%%%%%%%%%%%%%%%%%%%%%%%%%%%%%%%%%%%

In the P band, we collected baseband data over a bandwidth of 64\,MHz. After coherently de-dispersing and folding, we could not detect any pulsations from \src. This can be ascribed to scattering of the radio signal by the interstellar medium. Indeed, evidence of scattering was seen in the lower part of the L band by other telescopes \citep{lower20,joshi20}, which report scattering timescales of $\tau_s\sim 44$\,ms at 1\,GHz and $\sim$500\,ms at 600\,MHz. These imply $\tau_s\sim$3.5\,s at 336\,MHz, hence pulsations in the P band are likely completely smeared out. %This is supported by the $\tau_s\sim$?.?\,s at 1548\,MHz determined for the SP in Fig.\,\ref{fig:radio}.  t

Finally, we searched for \src\ in archival Parkes data. We found one observation at offset of $2.9'$ taken on 1998 August 01 at 1374\,MHz.
%as part of the PMSurvey \citep{manchester01} Essential? 
No pulsations down to a S/N $=7$ were detected, implying an upper limit to the flux density of 0.12\,mJy.   

%%%%%%%%%%%%%%%%%%%%%%%%%%%%%%%%%%%%%%%%%%%%%%%%%%%%%%%%%%%%%%%%%%%%%%%%
\begin{figure*}[h]
\centering
%\resizebox{\hsize}{!}{
\includegraphics[angle=0,width=13cm]{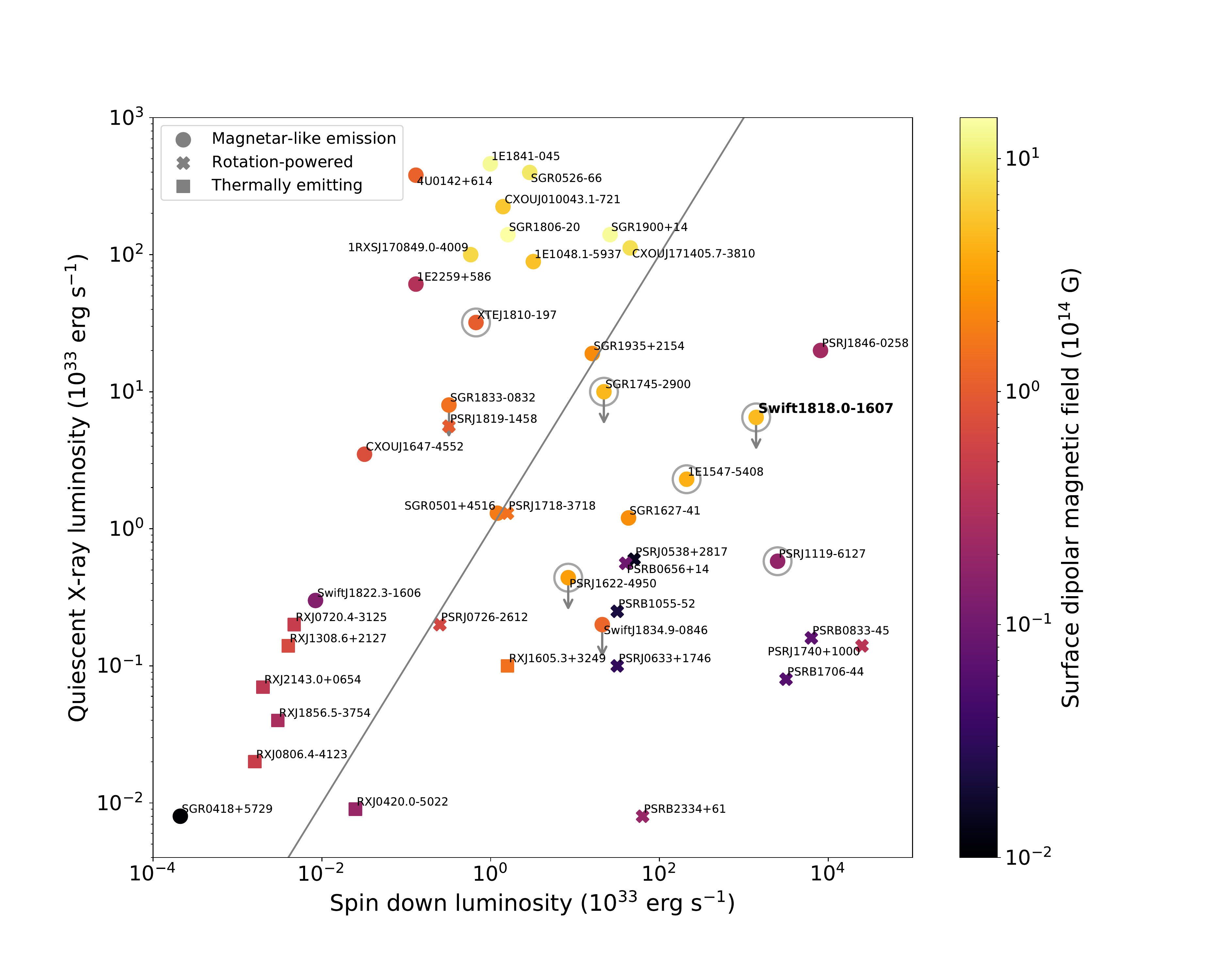}
\includegraphics[angle=0,width=13cm]{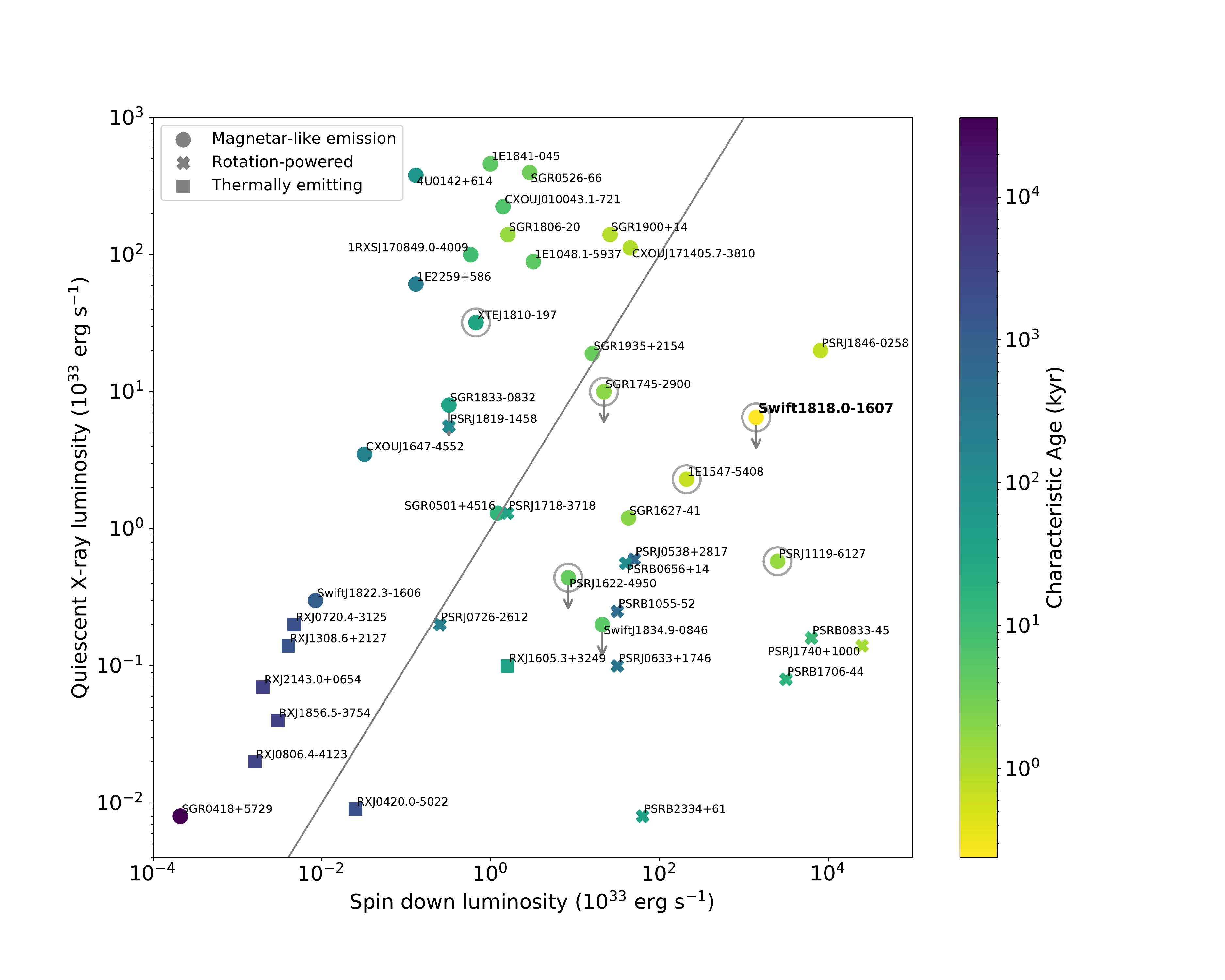}
%}
%\vspace{-6cm}
\caption{\label{lumvsedot} Quiescent X-ray luminosity as a function of the spin-down power for different classes of 
isolated X-ray pulsars, including \src\ (in bold). Circles denote radio-loud magnetars. The gray line indicates the equality line for the two parameters. Markers are color-coded according to the strength of the dipolar magnetic field at the pole (\emph{top}) and the characteristic age (\emph{bottom}). Values are from the Magnetar Outburst Online Catalogue (\url{http://magnetars.ice.csic.es/}; \citealt{cotizelati18}), with updates for \psr\ and \sgra\, \citep{camilo18,ybk17}.}
\end{figure*}
%%%%%%%%%%%%%%%%%%%%%%%%%%%%%%%%%%%%%%%%%%%%%%%%%%%%%%%%%%%%%%%%%%%%%%%%

\section{Discussion}
\label{discussion}

With a spin period of 1.36\,s, \src\ is among the fastest magnetars, in between the very active magnetar 1E\,1547.0--5408 (2.1\,s, also a radio emitting one; \citealt{camilo07}) and the allegedly rotation-powered pulsars PSR\,J1846--0258 (0.33\,s) and
PSR\,J1119--6127 (0.41\,s), which underwent magnetar-like
outbursts \citep{gavriil08,archibald16}. 

Our X-ray timing measurements of \src\ can be used to infer: (i) the characteristic age
\mbox{$\tau_c=P/(2\dot{P})\simeq240$\,yr}; 
%which is within $10$\,\% of the real age if the initial period is $P \lesssim 100 $\,ms, and if the dipolar magnetic field has remained constant since birth; 
(ii) the spin-down luminosity $\dot{E}_{\mathrm{rot}}=4\pi^2 I
\dot{P}P^{-3}\simeq1.4\times10^{36}$\,\lum, assuming a moment of inertia $I\approx10^{45}$\,g\,cm$^2$; (iii) the intensity of the dipolar component of the magnetic field at the pole,  $B\approx6.4\times10^{19}(P\dot{P})^{1/2}\simeq7\times10^{14}$\,G using the classical formula for an orthogonal rotator in vacuum. These values are compatible with the results from radio timing \citep{champion20}.
With $\tau_{\mathrm{c}}=240$\,yr, \src\ possibly represents the youngest pulsar discovered to date in the Galaxy, seconded by PSR\,J1846--0258. However, we caution that the age of \src, apart from the uncertainties connected to similar estimates, needs confirmation by a $\dot{P}$ measurement during quiescence and by the detection of its supernova remnant.

The observed 9\,ms burst with average luminosity $L_{\rm burst}\sim2\times 10^{39}d_{4.8}^2$\,\lum\ (\S\ref{results}), and a persistent X-ray spectrum at the outburst peak modeled by a blackbody of $\sim$1\,keV and a dim non-thermal component, are commonly seen during magnetar outbursts \citep[e.g.][]{cotizelati18}.
The radio emission of \src\, is not dissimilar to what observed in other radio magnetars. The period measured with SRT in radio is compatible within 3$\sigma$ with our X-ray timing parameters (\S\ref{section:timing}).
%$W50$ is smaller than determined at comparable central frequency in \citep{champion20}, possibly reflecting the variability in profile shape and flux density often showed by magnetars.
The fluence distribution of the strongest single pulses can also be fit with a power-law having index $-3.7\pm 0.3$ (1$\sigma$ uncertainty, Fig.\,\ref{fig:radio}). This, together with the fact that $W50_{\rm SP} \ll W50$, and with the high values of $R_{\rm E}$, are all reminiscent of the giant pulses observed in dozens of radio pulsars \citep[e.g.][]{oronsaye15}.

The ratio $S_{\rm SP}/S_{\rm ave}\sim 0.50$, where $S_{\rm SP}$ is the average flux density associated to the sum of the 53 single pulses, is independent of the uncertain flux calibration (\S\ref{section:srt}), and implies that at least 50\% of the total energy of the radio emission from \src\ is released in a form which resembles that of giant pulses. They have an average cadence of $\sim$0.9\,min$^{-1}$ (one burst every $\sim$50 rotations) and fluence larger than 1.3\,Jy\,ms. This means that, at the time of the SRT observation, the underlying radio emission mechanisms of \src\ were dominated by the sporadic emission of a succession of strong single radio bursts (with a flux density 1--2 orders of magnitude larger than the average flux density of the remaining pulsed emission), in contrast to what is typically seen in ordinary radio pulsars. Interestingly, if the source had been located 2--3 times farther, the regular pulsed emission could have been completely undetectable and the sporadic strong pulses interpreted as emitted by a rotating radio transient (RRAT; \citealt{mclaughlin06}), for some of which a link with the magnetar population has been proposed \citep{rea09,lyne09}. According to observations taken 12 days later at Parkes \citep{lower20}, the SRT pointing could have captured a transient behavior of \src, in agreement with the highly variable phenomenology shown also by other radio magnetars. 

%Complementarily, the X-ray luminosity provides precious information. 
Fig.\,\ref{lumvsedot} shows the quiescent X-ray luminosity (estimated as explained by \citealt{cotizelati18}), as a function of the spin-down luminosity for all neutron stars (NSs) that showed
magnetar-like emission, some high-$B$ radio pulsars
with detected X-ray emission, and the isolated thermally emitting
NSs \citep{turolla09}. This figure shows that the balance
between magnetic energy (related to the quiescent luminosity) and
rotational power might differ considerably between different sources of the same class.
Most of the radio magnetars can count on large rotational energy \citep{rea12}, and this is also the case for \src . 
The X-ray non-thermal luminosity in quiescence expected from the empirical $L_{\mathrm{X}}$--$\dot{E}_{\mathrm{rot}}$
relation for rotation-powered X-ray pulsars by \citet[][]{shibata16} is
$L_{\mathrm{X}}=3^{+2}_{-1}\times10^{32}$\,\lum\ (0.5--10\,keV). This value is consistent with the non-detection in
the archival data (Table\,\ref{tab:observations}), but much smaller
than the outburst value,
$8\times10^{34}d^2_{4.8}$\,\lum\ (\S\ref{xrayanalysis}). The X-ray conversion factor is $L_{\mathrm{X}}/\dot{E}_{\mathrm{rot}}<1$ in quiescence, and even at the outburst peak.

NSs with true age of a few centuries are expected to be still hot, with thermal luminosity normally exceeding $10^{34}$\,\lum\ \citep{pons09,vigano13}. Moreover, when high magnetic fields are taken into account, the Joule dissipation of the currents in the crust keeps the surface even hotter. According to crustal-confined magnetic field evolutionary models \citep{vigano13}, we should expect a minimum quiescent thermal luminosity of at least
$L_{\rm qui} = (5$--$7)\times10^{34}$\,\lum, or even higher if the NS has an envelope of light-elements, or the magnetic field has additional small scales components and/or a toroidal component. 

The value we derive for the quiescent luminosity of \src\, is $<10^{34}d^2_{4.8}$\,\lum\ (see Fig.\,\ref{lumvsedot}), rather low given the magnetic field ($B\lesssim 10^{14}$\,G) of this young magnetar. This can be explained if: a) the current $\dot{P}$ value is higher than in quiescence due to the extra-torque that might act during the outburst, as observed in other magnetars (e.g., \citealt{livingstone11}), thus implying that the real age is actually larger than 240 yr; b) currents are living only in the core, there are no toroidal components, and the NS underwent a fast cooling phase (meaning direct URCA processes and/or early superfluid transition; e.g. \citealt{page11}); c) the source is farther than estimated from the DM. In this respect, we note that distances inferred from the DM have large uncertainties for individual objects, and in the case of \src, the value also strongly depends on the model for the Galactic electron density, 4.8\,kpc using the YMW16 model \citep{yao17} and 8.1\,kpc with the NE2001  \citep{cordes02}. 

Our future observations of the evolution of the source towards quiescence will help to address  whether the diffuse emission surrounding the source is a dust scattering halo due to the bursting activity or the brightening of \src\ \citep[see][]{tiengo10}, as well as to constrain the quiescent spin-down rate. 

Overall, we see the emission observed from \src\ as another example of the possible ubiquitous presence of magnetar-like activity in pulsars of any class.

\acknowledgments

%The first four authors contributed equally to this work. 
This research is based on observations with \xmm\ (ESA/NASA), \nustar\ (CaltTech/NASA/JPL), \swift\ (NASA/ASI/UKSA), and on  data retrieved through the CXC and the NASA/GSFC’s HEASARC archives.
We acknowledge the support of the PHAROS COST Action (CA16214). NR, AB, DV, CD, MR, AGG and VG are supported by the ERC Consolidator Grant ``MAGNESIA" (nr.817661) and acknowledge funding from grants SGR2017-1383 and PGC2018-095512-BI00. AB and FCZ are also supported by a Juan de la Cierva fellowship. 
GLI, SM, AT, and RT acknowledge financial support from the Italian MIUR through PRIN grant 2017LJ39LM. 
AP, AR, MB and LS acknowledge funding from the grant ``iPeska'' (INAF PRIN-SKA/CTA; PI Possenti). LS acknowledges funding from ASI-INAF agreements 2017-14-H.O and  I/037/12/0.
\facilities{\xmm\ (EPIC), \swift\ (BAT,XRT), \nustar, SRT}
\software{SAS \citep{gabriel04}, FTOOLS \citep{blackburn95}, XSPEC \citep{arnaud96}, NuSTARDAS, PRESTO}
 
\bibliographystyle{aasjournal}
%\bibliography{biblio}

\end{document}